\documentclass[prb,showpacs,showkeys,preprintnumbers,superscriptaddress,a4paper,twocolumn]{revtex4}
\usepackage[T1]{fontenc}
\usepackage[utf8]{inputenc}
\usepackage{graphicx}
\usepackage{lmodern}
\usepackage[english]{babel}
\usepackage{amsmath,amssymb}
\usepackage{amstext}
\usepackage[colorlinks=true,linkcolor=red]{hyperref}
\usepackage{color}														%	remove from the final version
\usepackage{graphicx}

\begin{document}

\title{Collective spin 1 singlet phase in high pressure oxygen}

\author{Y. Crespo}
\affiliation{The Abdus Salam ICTP, Strada Costiera 11, I-34151 Trieste, Italy}
%\address{The Abdus Salam ICTP, Strada Costiera 11, I-34151 Trieste, Italy}
\author{M. Fabrizio}
\affiliation{International School for Advanced Studies (SISSA), and CNR-IOM Democritos, Via Bonomea 265, I-34136 Trieste, Italy}
\author{S. Scandolo}
\affiliation{The Abdus Salam ICTP, Strada Costiera 11, I-34151 Trieste, Italy}
\author{E. Tosatti}
\affiliation{The Abdus Salam ICTP, Strada Costiera 11, I-34151 Trieste, Italy}
\affiliation{International School for Advanced Studies (SISSA), and CNR-IOM Democritos, Via Bonomea 265, I-34136 Trieste, Italy}
\includegraphics[width=17.0cm,height=9.2cm,angle=0]{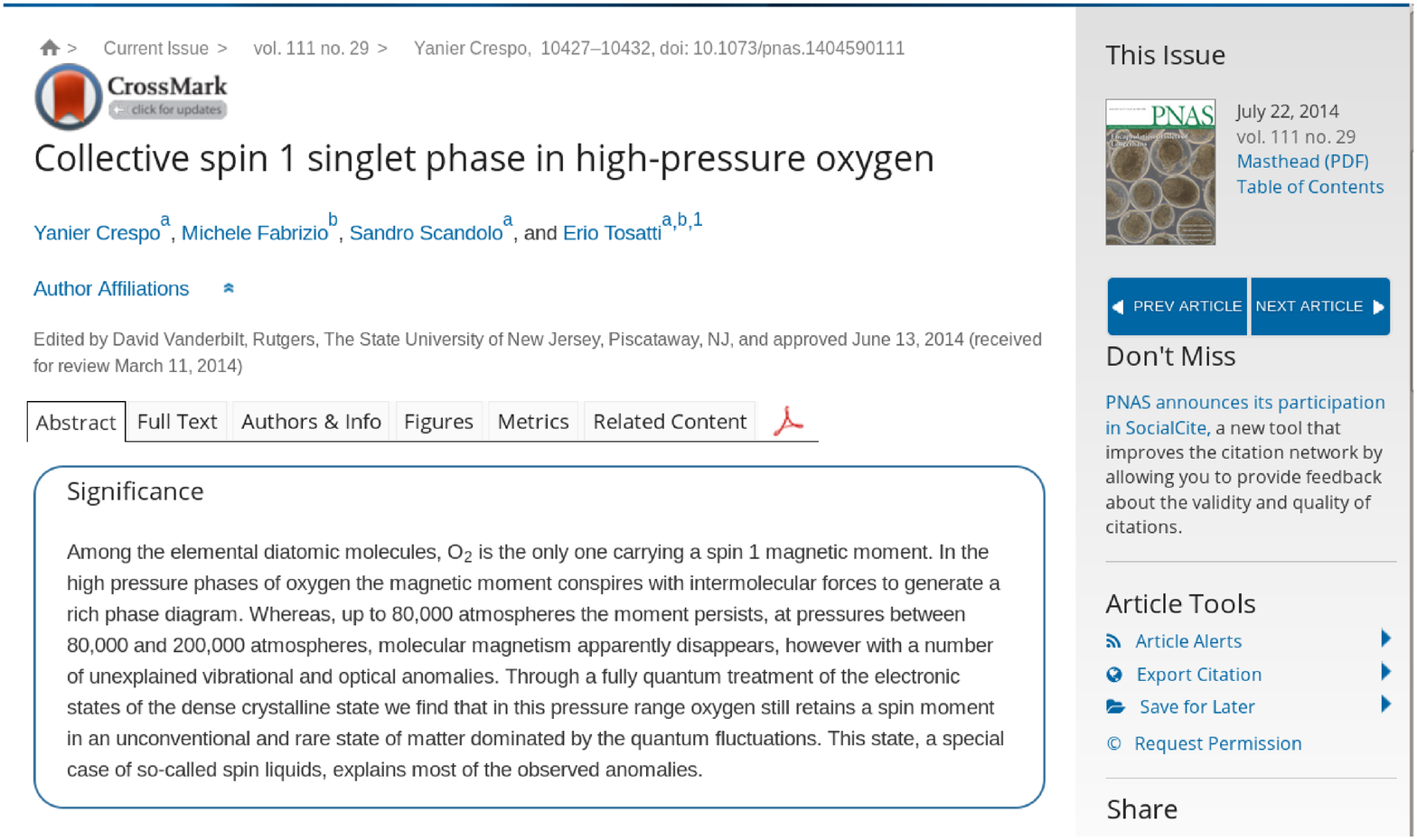}
% Ali A. Hassanali\affil{1}{Department of Chemistry and  Applied
% Biosciences, ETH Zurich and Universit\`{a} della Svizzera Italiana,via
% G. Buffi 13, CH-6900 Lugano, Switzerland}, Federico Giberti\affil{1},
% J\'{e}r\^{o}me Cuny\affil{2}{Laboratoire de Chimie et Physique Quantiques - UMR 5626,
% Toulouse, France.}, Thomas K\"uhne\affil{3}{Institute for Physical Chemistry, 
% University of Mainz, Mainz, Germany} 
% and Michele Parrinello$^{*}$}

% \date{\today}

\begin{abstract}

{\it This work has just been published in PNAS, {\bf 111}, 10427 (2014)}: Oxygen, one of the most common and important elements in nature, has an exceedingly well explored phase 
diagram under pressure, up and beyond 100 GPa. At low temperatures, the low pressures antiferromagnetic phases below 8 GPa where O$_2$ molecules
have spin  S=1 are followed by the broad apparently nonmagnetic $\epsilon$ phase from about 8 to 96 GPa. In this phase which is our 
focus molecules group structurally together to form quartets while switching, as believed by most, to spin S=0.  Here we present theoretical results
strongly connecting with existing vibrational and optical evidence, showing that this is true only above 20 GPa, whereas the S=1 molecular state 
survives up to at about 20 GPa.  The  $\epsilon$ phase thus breaks up into two: a spinless $\epsilon_0$ (20-96 GPa), and another $\epsilon_1$ 
(8-20 GPa) where the molecules have  S=1 but possess only short range antiferromagnetic correlations. A local spin liquid-like singlet ground state akin 
to some earlier proposals and whose optical signature we identify in existing data, is proposed for this phase. Our proposed phase diagram thus has a first order 
phase transition just above 20 GPa, extending at finite temperature and most likely terminating into a crossover with a critical point near 30 GPa and 200 K.

\end{abstract}

\keywords{epsilon oxygen | high pressure  | phase transition | spin singlets }

% \abbreviations{}

\maketitle

% \section{Significance Statement}
% 
% Among the simple diatomic molecules, O$_2$ is the only one carrying a spin one magnetic 
% moment. In the condensed phases of elemental oxygen the molecular 
% magnetic moment conspires with intermolecular forces to generate an 
% unusually rich phase diagram. Whereas up to pressures of 80,000 atmospheres
% there is evidence that the magnetic moment persists, at pressures between 80,000 
% and 200,000 atmospheres molecular magnetism apparently disappears, and does so with a 
% number of unexplained vibrational and optical anomalies. Here we develop 
% a fully quantum treatment of the electronic states of the dense crystalline state 
% and find that in that pressure range oxygen actually retains a spin moment in an 
% unconventional and  rare state of matter dominated by the quantum fluctuations 
% of the spins. Such a state, a special case of so-called spin liquids, explains most 
% of the observed anomalies. \\

%end of significance statement 
%  \begin{figure*}[!b]% 
% 		\centering
% 		\includegraphics[width=12.0cm,angle=0]{PNAS.eps}
% 		\caption{}
% \end{figure*}
% \newpage
\section{Introduction}

\begin{figure}[!t]% 
		\centering
		\includegraphics[width=8.0cm,height=5.2cm,angle=0]{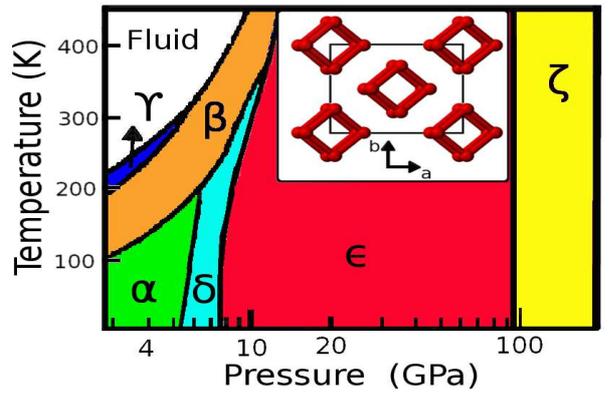}
		\caption{\label{fig:PDiag}{(Color online) Currently accepted phase diagram of oxygen (adapted from Ref.\cite{Freiman2004}). The inset shows the unit cell of $\epsilon$-O$_2$ ~\cite{Loubeyre2006,Fujihisa2006}}}
\end{figure}

Molecular systems display at high pressure a horn of plenty of intriguing phases. That is especially true of molecular 
oxygen, whose diatomic molecule survives unbroken up to at least 133 GPa~\cite{Weck_prl_2009}, and where the original spin S=1 of the gas phase plays 
an important role.  In the phase diagram of O$_2$ (Fig.\ref{fig:PDiag}) we focus on the wide $\epsilon$-O$_2$ phase between 8 and 
96 GPa, a phase which has long intrigued the community~\cite{Freiman2004}. Unlike the two bordering phases, 
$\delta$-O$_2$ an antiferromagnetic (AF) S=1 correlated insulator at lower pressure, and $\zeta$-O$_2$ a 
regular and superconducting nonmagnetic (NM) 
metal at higher pressure, $\epsilon$-O$_2$ is an insulator of more complex 
nature.  
Structurally, 
high-pressure X-ray diffraction~\cite{Loubeyre2006,Fujihisa2006} revealed in the last decade that 
at the $\delta - \epsilon $ transition at P $\approx$ 8 GPa  the close-packed O$_2$  planes undergo a large distortion giving 
rise to molecular O$_8$ ``quartets'',  (inset in Fig.\ref{fig:PDiag}). 
Spin-polarized neutron diffraction showed that simultaneously there is a collapse of  
long-range AF N\'eel order at the $\delta-\epsilon$ transition ~\cite{GoncharenkoPRL2005}.  That 
observation unfortunately 
did not provide conclusive information about the nature of the ground state in $\epsilon$-O$_2$ and in particular about 
any further role played in $\epsilon$-O$_2$ by the spin of individual molecules if any. 
It has been tempting to imagine that the O$_2$ molecular magnetic state could simply %switch 
collapse
from S=1 to S=0 at the $\delta - \epsilon $ transition.
In support of this idea it can be noted that the metallic state band structure
of a hypothetical undistorted nonmagnetic O$_2$ ~\cite{Serra1998} is prone to turn spontaneously insulating through
a Peierls type distortion, for example dimerizing ~\cite{PRL.Gorelli.99, Neaton_Ashcroft} , or tetramerizing~\cite{Hemley.Militzer.2006} the molecules.
Further density functional theory (DFT) 
calculations strengthened that picture, showing the quartet distorted geometry~\cite{Oganov_PRB_2007} 
drives the undistorted metal to a band insulator. Moreover, DFT calculations showed that this state 
exhibits O$_2$ vibrations whose frequency and pressure evolution are, between 20 and 96 GPa, in good  agreement 
with infrared (IR) and Raman data ~\cite{Scandolo_SSC} 
 (see Fig.\ref{fig:MIR}(a) ).    
 
An alternative, much more speculative picture insisting instead on a correlated state where O$_2$ retains its S=1 gas phase spin, and where the quartet of AF 
coupled sites yields a singlet ground state as the basic O$_8$ block of $\epsilon$-O$_2$,
~\cite{Gomonay_Loktev} provided no such successful predictions and had limited following so far -- see however Ref.\cite{Bartolomei11}. 
 
Let us provide first some  background before embarking in theoretical calculations.  
Several experimental authors \cite{PRL.Gorelli.99,PhysB.Gorelli.99,Gorelli-PRB-2001,Akahama.1996,Carter.1991} had noted 
that the behaviour of O$_2$ in the 8-20 GPa 
pressure range is in many ways anomalous compared with that above 20 GPa ~\cite{Ulivi_book} . Focusing on  
O$_2$ vibrations, the good agreement found by Pham \emph{et al.} ~\cite{Scandolo_SSC} 
between the band insulator calculations and the measured IR and Raman frequencies (Fig.\ref{fig:MIR}(a)) is  limited to above P $> 20$ GPa. 
Below this pressure, the IR mode reverts non-monotonically upwards approaching the high frequency Raman mode 
at the $\delta$-O$_2$ phase boundary of 8 GPa, rather than  dropping steadily as predicted by the band insulator. 
Subtly, but not less significantly, the  O$_2$ Raman data (Fig.\ref{fig:MIR}(a)) show that 20 GPa marks a delicate but definite breaking point 
with a lower rate of decrease of the mode frequency with decreasing pressure. Both IR and Raman 
elements hint at a possible switch of individual O$_2$ molecules from S = 0 to S = 1 upon decreasing pressure near 20 GPa. 
In particular, the frequency 
gap between the IR  mode, where nearest neighbour O$_2$ vibrate out of phase, and the Raman mode where 
they vibrate in phase (see insets in Fig.\ref{fig:MIR}(a)) is proportional to the IR effective charge, connected 
with electron current ``pumping'' between an instantaneously extended molecule (which attracts electrons)  and a neighboring 
compressed one (which expels them) in the IR 
mode. This electron 
current is absent in the Raman mode, where neighboring molecules vibrate in phase. If the molecules have spin and 
correlations are strong, the electron hopping is reduced and the current magnitude, proportional to the IR effective charge, 
must drop compared to the 
nonmagnetic band state. Thus the onset of molecular spin upon decreasing pressure should be accompanied by a collapse 
of the IR intensity and of the IR -Raman splitting, as is indeed observed (see Inset in Fig.\ref{fig:Intensities}). 
With this background, experimental facts suggest dividing the vast $\epsilon$-O$_2$ phase into two. A higher pressure phase
between 20 and 96 GPa, which we may call $\epsilon_0$-O$_2$, where there is no molecular spin and whose physics 
including lattice vibrations is well described as a band insulator whose gap is due to the Peierls-like quartet distortion,   
and a lower pressure phase between 8 and 20 GPa
which we shall call $\epsilon_1$-O$_2$ where that picture fails, and molecular spin probably resurrects signaling that strong correlations 
coexist within the quartet distortion of the molecules. 
The long range N{\'e}el order typical of the lower pressure undistorted phases is absent here, suggesting that it might be replaced by some kind of 
correlated singlet state such as that of  Ref.\cite{Gomonay_Loktev}. However, strong and verifiable predictions
about the same measured vibrational and optical that are well described by the band insulator in  $\epsilon_0$-O$_2$ have 
not been advanced for this type of state.

Our work has the following aims: (i)  Provide first principles quantitative calculations of electronic and vibrational properties of same quality for 
both $\epsilon_0$-O$_2$ and $\epsilon_1$-O$_2$,  including Raman and IR vibrational frequencies and intensities that 
could be compared with experiment in both regimes. The approach should allow, even if at the approximate mean field level, 
for the presence of molecular spin whenever that should lower the total enthalpy; (ii) Describe the nature, stability, and
optical properties of the resulting model S=1  collective singlet state description of $\epsilon_1$-O$_2$, extending 
that in literature~\cite{Gomonay_Loktev, Bartolomei11}, particularly including quantum fluctuations and ; (iii) Predict 
the new phase diagram displaying 
a first order $\epsilon_1- \epsilon_0$ low temperature phase transition near 20 GPa, and a novel phase line predicting
at high temperature a new critical point roughly at 30 GPa and 200 K inside the broad $\epsilon$-O$_2$ phase.
   
\section{First Principles Calculations and Results}   

\begin{figure}[!t]% 
		\centering
 		\includegraphics[width=3.2in,angle=0]{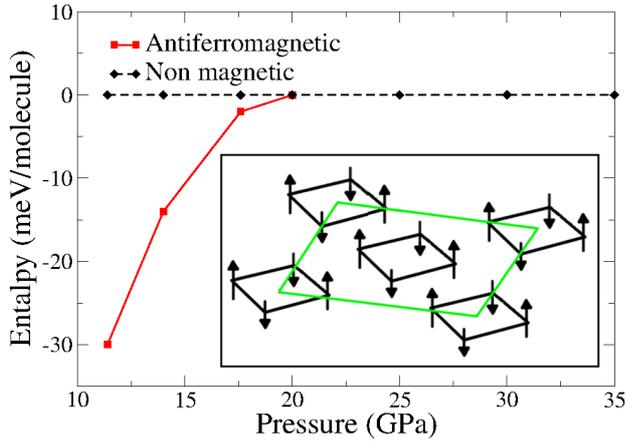}
	        \caption{\label{fig:Entalphy}{ Enthalpy difference calculated by DFT+$U$ between non-magnetic and 
		 antiferromagnetic states in $\epsilon$-O$_2$.  Inset: antiferromagnetic 
		configuration inside the O$_8$ ``quartets''.}}
\end{figure}  

      \begin{figure}[!t]% 		
		\centering
 		\includegraphics[width=8.0cm,height=9.2cm,angle=0]{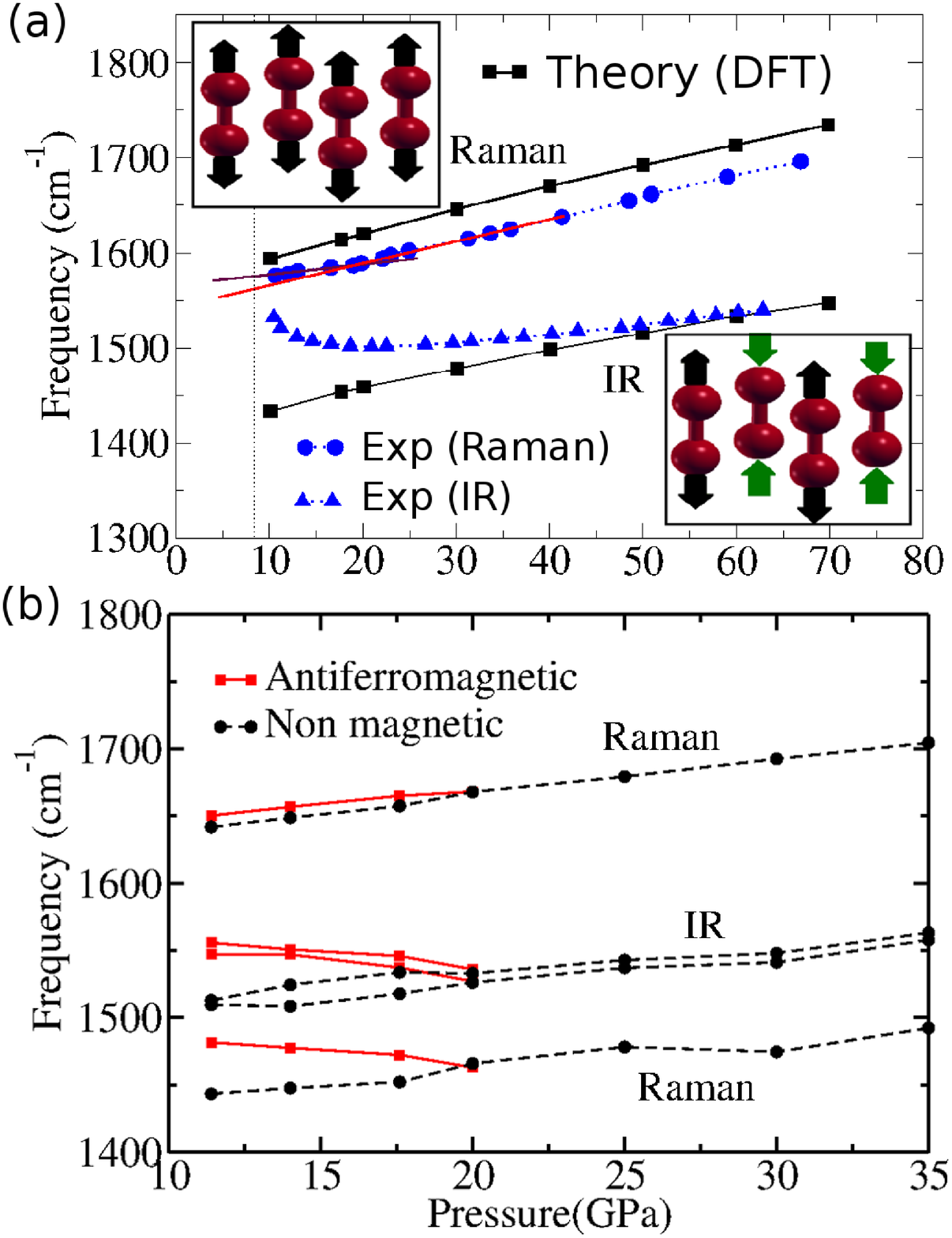}
		\caption{\label{fig:MIR}{(Color online) Vibrational frequencies of the O$_2$ stretching modes as a function of pressure.
		~(a) Solid line with squares: calculated Raman and IR modes 
 		from Ref.\cite{Scandolo_SSC}).  Dashed line with circles: Raman mode data from Ref.\cite{Akahama.1996}. Dashed line 
                with triangles: IR mode data from Ref.\cite{PRL.Gorelli.99}. 
		Displacement patterns for the IR  and Raman modes are shown in the lower right and upper left insets, respectively.  
		(b) Calculated Raman and IR modes for the non-magnetic and the antiferromagnetic configurations (this work).
		}}
	\end{figure} 
	
Our starting point is a density-functional theory (DFT) electronic structure calculation with full structural optimization 
for the whole $\epsilon$-O$_2$ pressure range. We employed spin-polarized, self-interaction corrected calculations (DFT+$U$)~\cite{AnisimovPRB.44,
AnisimovPRB.48,LiechtensteinPRB.52} as implemented in Quantum-Espresso \cite{qe1}. By allowing for static magnetic 
polarization we permit the possible presence of molecular spin to emerge -- of course at the price of assuming it to exist
in static and thus symmetry-breaking form.  The GGA exchange-correlation functional 
was employed in the version of Perdew, Burke and Ernzerhof \cite{pbe}, and DFT+{\it{U}} calculations were carried out in the 
simplified version of Dudarev {\it {et al.}} \cite{Dudarev.1998} as implemented in the Quantum-Espresso code \cite{cococcioni}. 
The inclusion of a Hubbard $U$ for oxygen p-states is called for to provide cancellation of self-interactions still present 
in simple generalize gradient approximation (GGA) DFT.  Whereas for the uncorrected choice $U$=0 the calculated band gap 
is unrealistically small and spin polarization does not arise,~\cite{Oganov_PRB_2007,Scandolo_SSC}  
calculations in a reasonable range of values of the parameter $U$ (0.8, 1.0, 1.5 and 2.0 eV) yield antiferromagnetism below 17.6, 20.0, 25.0 
and 30.0 GPa respectively. We selected {\it {$U$=1.0}} eV, as the value yielding the transition pressure that fits more reasonably 
the experimental findings to be described below in the vibrational spectra.~\cite{PRL.Gorelli.99,PhysB.Gorelli.99,Akahama.1996}. 
Their frequency and intensity behaviour does not change with the value of $U$. 
We note that this method to reduce self-interactions is convenient but by no means unique, and other possibilities such as
hybrid functional approximations could have been adopted.
A $4\times 4\times 4$ Monkhorst-Pack k-point mesh~\cite{k-point-mesh}  Brillouin zone sampling was used throughout. 
The crystal structure (C2/m ) and the initial atomic positions were taken from the experimental data \cite{Fujihisa2006}. Lattice and internal parameters 
were then fully optimized at different pressures until forces were smaller than 10$^{-5}$ a.u. The vibrational spectra were calculated 
using the {\it {fropho}} code that calculates phonons based on the Parlinski-Li-Kawazoe method \cite{fropho}. Infrared intensities 
were calculated by density functional perturbation theory \cite{Phonons.QE} by single-point DFT calculations with the fully DFT+$U$ relaxed 
crystal structures. 

		 \begin{figure}[!t]% 		
		\centering
 		\includegraphics[width=8.0cm,angle=0]{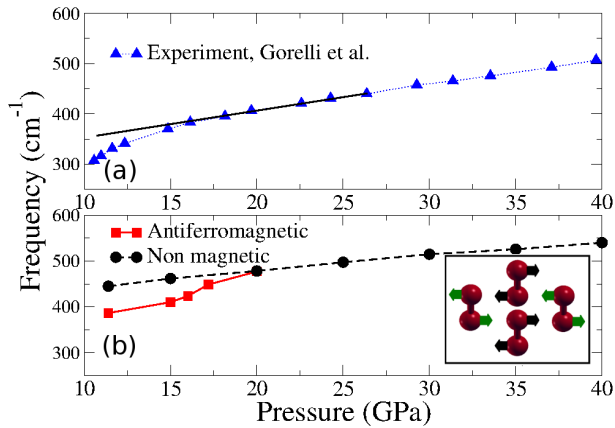}
		\caption{\label{fig:FIR}{Frequency of the far-IR vibrational mode as a function of pressure. (a) Data from Ref.\cite{PRL.Gorelli.99} 
               (b) Calculations using the non magnetic and the antiferromagnetic configurations. Inset: calculated displacement pattern of the far-IR mode.}}
		\end{figure}

The optimized structure has O$_2$ molecules forming quartets in each plane, quite close to the experimental
ones \cite{Fujihisa2006}.  A clear AF state prevails below 20 GPa, as shown by the enthalpy difference of Fig.\ref{fig:Entalphy} 
between the NM and the AF state, where molecules are simultaneously quartet distorted and antiferromagnetically spin polarized.
As anticipated, the predicted low pressure resurgence of molecular spin comes with an unrealistic AF static long range order which is absent
in experiment \cite{GoncharenkoPRL2005}. How this artificial mean-field symmetry breaking can be removed by quantum fluctuations 
will be described later. Ignoring that for the moment, and assuming the mean-field to yield as usual a reasonable total energy, 
we can exploit our first principles calculations to obtain 
a prediction for the change of other properties brought about by the instantaneous presence of molecular spin. 
We calculated, with the DFT+$U$ approach, the vibrational spectra of both the nonmagnetic and magnetic states of $\epsilon$-O$_2$.
As is shown in Fig.\ref{fig:MIR}(b) the onset of spin breaks the monotonic drop of both Raman and IR
mode frequencies with decreasing pressure, that was predicted for the nonmagnetic state by previous calculations \cite{Scandolo_SSC}
but which did not agree with experiment below 20 GPa. Direct comparison of Fig.\ref{fig:MIR}(b) with Fig.\ref{fig:MIR}(a) shows now a better 
agreement, confirming that both the non-monotonic rise of the IR mode and the slight stiffening of the Raman mode are spin-related.
Many other phonon modes are also influenced by the onset of spin. 

             \begin{figure}[!t] % fig 3
		\centering
 		\includegraphics[width=8.0cm,angle=0]{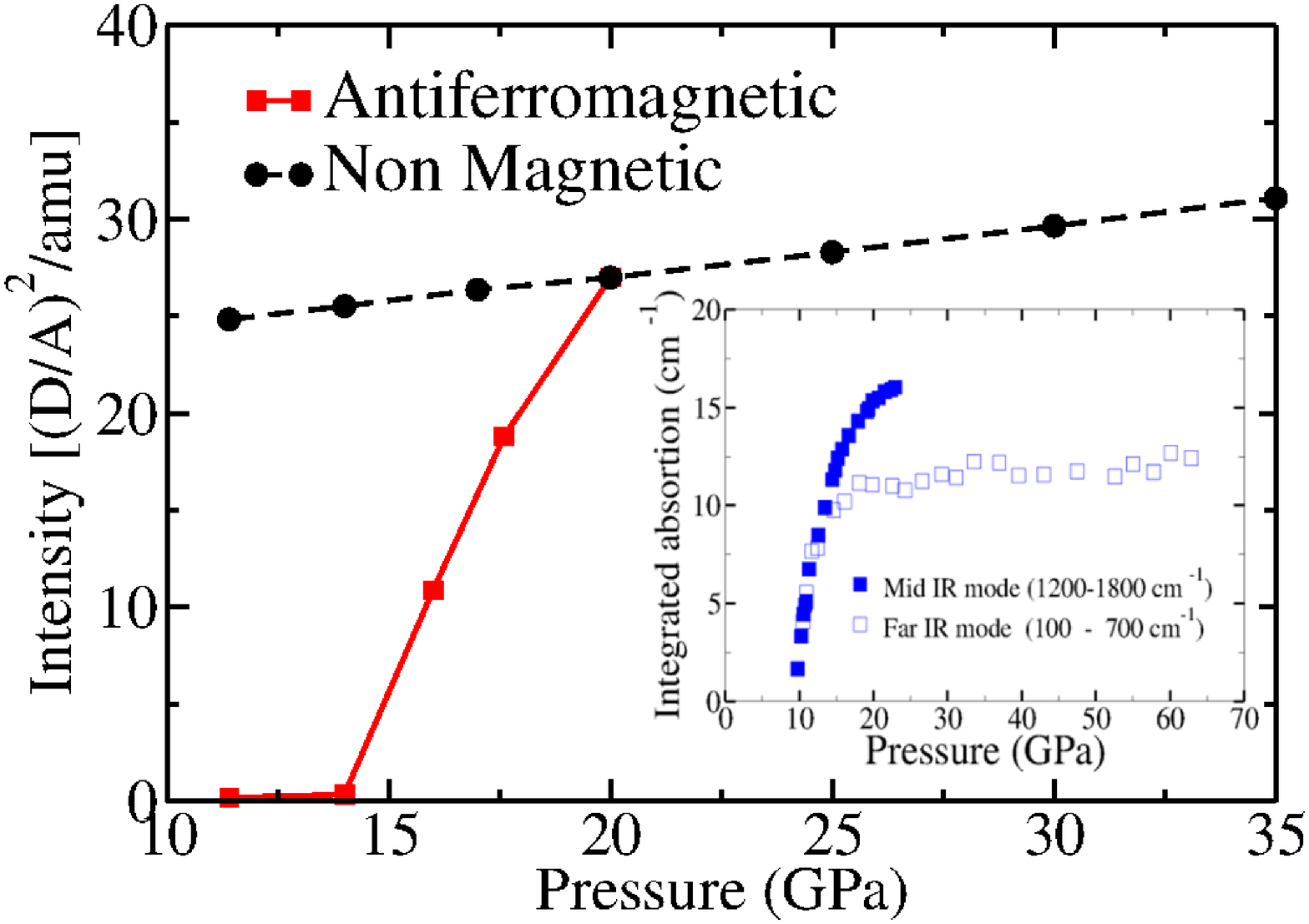}
		\caption{\label{fig:Intensities}{(Color online) Calculated Intensities for mid-IR modes (1500-1600 cm$^{-1}$) of both 
		non-magnetic and antiferromagnetic configurations (this work). Inset: solid squares, measured IR intensities of the mid-IR mode 
		(1200-1800 cm$^{-1}$); open squares,  far IR mode (200-600 cm$^{-1}$) intensities from Ref.\cite{PRL.Gorelli.99}}}
	      \end{figure}
	      
In Fig.\ref{fig:FIR} our calculated evolution of the main far infrared vibrational mode (see Fig.\ref{fig:FIR} (b)),  is shown to drop below 
20 GPa in agreement with experimental data (see Fig.\ref{fig:FIR} (a)). This evolution of both high and low frequency IR modes upon 
lowering pressure below 20 GPa  goes together with a corresponding change -- in fact a decrease -- of the mode effective charge. 
Fig.\ref{fig:Intensities} shows that the dramatic drop of the IR intensity observed in this regime (see Inset in Fig.\ref{fig:Intensities}), 
so far unexplained, is now well accounted for by the onset of molecular spin.

We already hinted at the main physical reasons why spin causes all these changes in the vibrational spectrum.  The first element is
that spin arises in connection with strong electron correlations, which characterize all lower pressure phases including $\delta$-O$_2$.
In the strongly correlated state, O$_2$ molecules reduce their mutual electron hopping, and tend to revert toward their gas phase state, 
which has spin $1$, with shorter bond length and about 71 cm$^{-1}$  higher vibrational frequency \cite{Herzberg_book66}. That explains why the highest 
(intramolecular) Raman mode reduces its softening rate below 20GPa (see Fig.\ref{fig:MIR}). The reduced IR effective charge, and the 
shrinking of the frequency gap between the out-of-phase IR and the in-phase Raman modes as calculated and observed implies a
reduced intermolecular electron hopping in the AF correlated state.  

For our subsequent understanding of the magnetic state below 20 GPa it is also important to estimate the effective Heisenberg exchange couplings among O$_2$ molecules.
Calling $J_1, J_2, J_3$ and $J_4$ the exchange values between the first, second, third and fourth neighbors (see Fig.\ref{fig:js.val})
we carried out constrained spin polarized DFT+$U$ calculations, based on the experimental structure at P = 11.4 GPa \cite{Loubeyre2006} and
a variety of six different AF configurations.  Fitting these results we obtained $J_1$ = 170.0 $\pm$ 30 meV, $ J_2$ = 35.5 $\pm$ 2 meV, 
$ J_3$ = 10.5 $\pm$ 2 meV $ J_4$ = 14.4 $\pm$ 4 meV. On account of the mean-field nature of the DFT calculations, these values 
are probably somewhat larger values than real, and should be considered upper bounds.

%%%% REVISED BY MICHELE
\section{ Lattice Singlet Model and Quantum Fluctuations of S=1 Spins}              
     
The previous section showed that the resurgence  of molecular spin below 20 GPa could simultaneously explain several observed
O$_2$ vibrational anomalies. However the mean-field long range AF order obtained by DFT disagreed with the lack of long range 
spin order found experimentally~\cite{GoncharenkoPRL2005}. A state where molecular spins are present without $T$ = 0  long-range order 
would constitute a kind of spin 1 liquid. In this section we discuss, extending the picture earlier proposed by Gomonay and 
Loktev~\cite{Gomonay_Loktev} and more recently pursued by Bartolomei \emph{et al.}~\cite{Bartolomei11} consisting
of an overall singlet state of an isolated quartet of $S=1$ molecules.

    \begin{figure}[!t]% 
		\centering
 		\includegraphics[width=8.0cm,,height=6.2cm,angle=0]{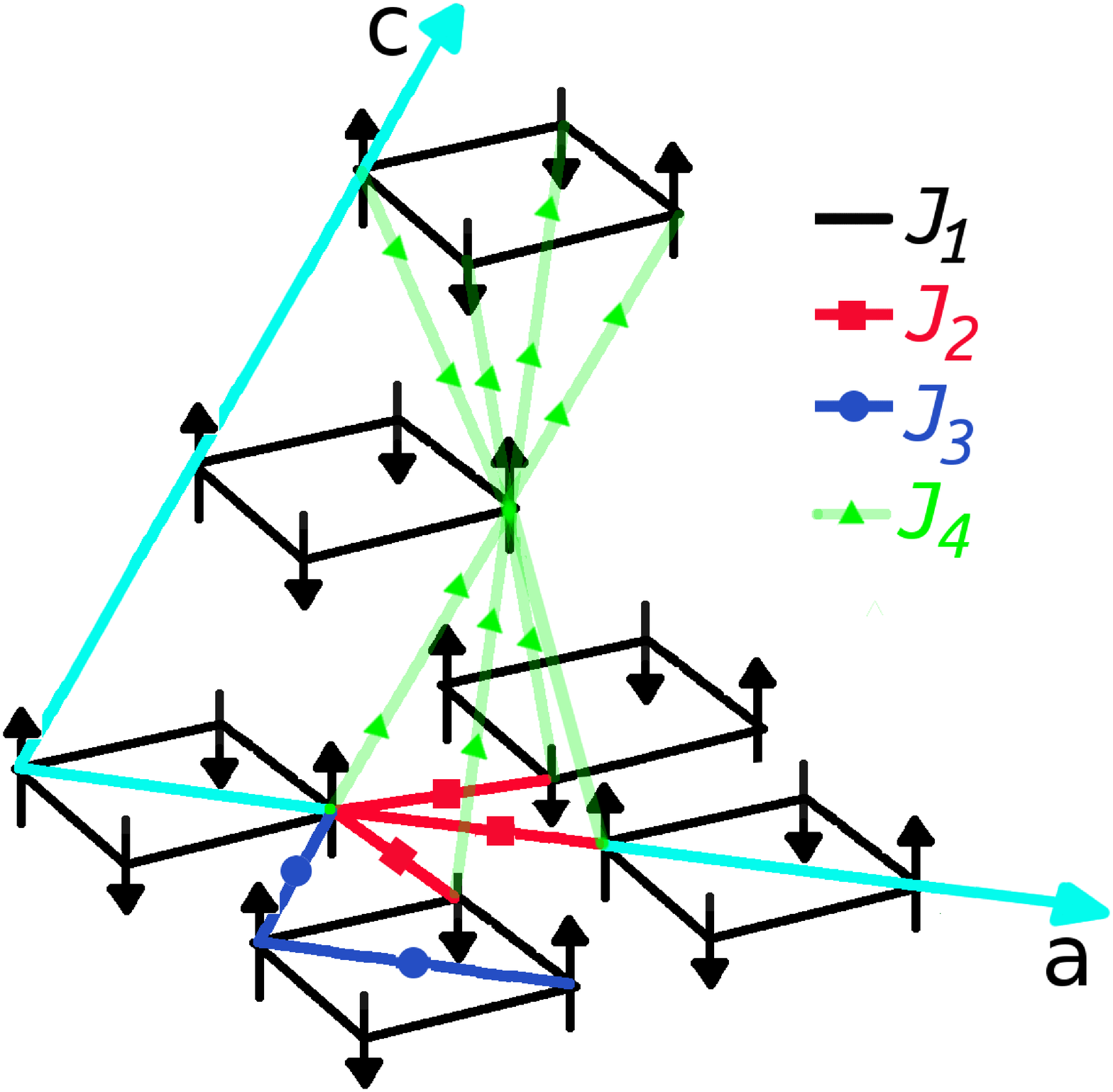}
		\caption{\label{fig:js.val}{(Color online) Long range antiferromagnetic order and interactions between the first 
		fourth neighbors $J_1$ (inside the quartet), $J_2$ with the two or three nearest quartets, $J_3$ third nearest 
		neighbors in the plane and $J_4$ with molecules located in the planes above and below.}}
     \end{figure}

\begin{figure}[!t]
\vspace{0.2cm}
\centerline{\includegraphics[width=7cm,height=6.2cm]{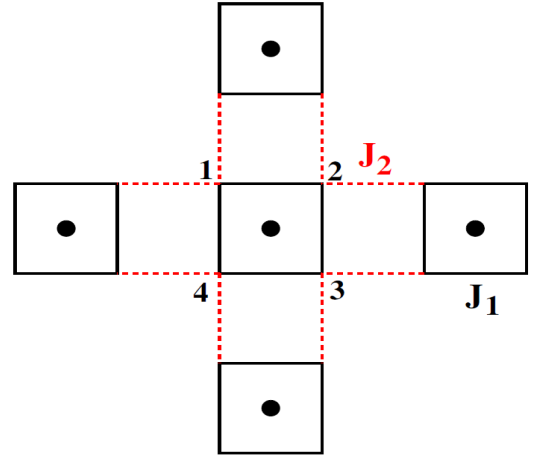}}
\caption{\label{Heisenberg_model}(Color online) Simplified two-dimensional Heisenberg model.}
\end{figure}

The quantum mechanical competition between AF long range order and an overall singlet state  must be pursued in the 
whole $\epsilon_1$-O$_2$ lattice.   
For that purpose we simplify 
the system as a 2D square lattice model made of plaquettes (quartets) of $S=1$ Heisenberg sites. 
Each site, representing an O$_2$ molecule, is  AF coupled to nearest neighbours within the same plaquette by 
AF exchange couplings $J_1>0$,  and to nearest neighbours in the next plaquettes by $J_2<J_1$, see Fig.\ref{Heisenberg_model}. 
Two different states compete: the Ne\'el  AF configuration, as obtained by DFT and which breaks spin SU(2) symmetry,  and a singlet, NM 
state that is akin to a collection of independent plaquettes, each in its singlet ground state.  
The singlet ground state of an isolated plaquette of energy $E_0=-6J_1$  is obtained by coupling second neighbor sites 1 and 3 
to $S_{13}=S_1+S_3=2$,  sites 2 and 4  to $S_{24}=S_2+S_4=2$, 
see Fig.\ref{Heisenberg_model},  and then coupling $S_{13}$ and $S_{24}$ to a total singlet $S= S_{13}+S_{24}=0$. 
The energy per site of an independent collection of plaquettes thus is (note that the number of plaquettes 
$N_{\,\square}$ is one quarter the number of sites)
\begin{equation}
 E_{\,\square} = - \frac{3}{2}\;J_1.\label{square}
\end{equation}
By comparison, the classical energy per site of the Ne\'el AF configuration is 
\begin{equation}
E_{\,\text{Ne\'el}} = -J_1 - J_2,\label{Neel}
\end{equation}
Therefore, for $J_2<J_1/2$ the non-magnetic collection of independent plaquettes 
will be lower in energy than the Ne\'el configuration.  If we use  $J_2\simeq 35$ meV and $J_1\simeq 170~\text{meV} > 2\,J_2$, 
we may conclude that the actual ground state is a collection of independent non-magnetic plaquettes.  
This state is akin to that proposed in Ref.~\cite{Gomonay_Loktev}.
%%%%
We observe further
that the  next-nearest-neighbor exchange $J_3>0$ frustrates and penalizes the Ne\'el configuration more than the 
non-magnetic one, leading to the energy balance  %For instance, an exchange $J_3$ 
within each plaquette from $E_{\, \square} - E_{\,\text{Ne\'el}} 
= -J_1/2+J_2\to -J_1/2 + J_2 - J_3/2$. Moreover, the inter-plane exchange $J_4$  gives no contribution to the classical energy,
since its effects cancel out as can be seen in Fig.\ref{fig:js.val}.
Since both $J_3$ and $J_4$ are anyhow smaller than $J_2$, we shall not consider them in the following  
analysis.
%%%%
    %%%%%%%%%%%%%%% FIG
%%%%%%%%%%%%%%%%%
The above treatment however is still 
crude, as it does not take into account quantum 
fluctuations. To evaluate their impact
we assess the stability of the non-magnetic state against quantum fluctuations. 
Elementary quantum fluctuations are built by combining into an overall singlet two separate spin fluctuations in  
neighboring plaquettes. The first excited state of the isolated plaquette is still obtained
by $S_{13}=2$ and $S_{24}=2$,  now coupled into a total spin $S=S_{13}+S_{24}=1$, at energy $J_1$ above the ground state. 
Let us denote as $|\mathbb{S}\rangle$ the plaquette singlet ground state, 
and  $|\mathbb{T},M\rangle$ the excited triplet with $S_z=M=-1,\dots,+1$. If we consider two nearest neighboring plaquettes, 
identified by the positions $\bf R$ and $\bf R'$, application of the exchange $J_2$ to the state in which both plaquettes 
are in the ground state 
\begin{eqnarray}
H_{J_2}\;|\mathbb{S};\bf R\rangle|\mathbb{S};\bf R'\rangle &=& 
-J_2\, \sum_{M=-1}^{+1}\,(-1)^M\, \times  \nonumber  \\
&& |\mathbb{T},M;\bf R\rangle|\mathbb{T},-M;\bf R'\rangle,
\end{eqnarray}
excites in both plaquettes the spin-triplet configurations, with the two coupled to a singlet. 
This process has an amplitude $\sqrt{3}\;J_2$. These two excitations could move, still remaining in a 
full singlet configuration, 
\begin{equation}
H_{J_2}\;|\mathbb{S};\bf R\rangle|\mathbb{T},M;\bf R'\rangle = 
-J_2\; |\mathbb{T},M;\bf R\rangle|\mathbb{S};\bf R'\rangle. 
\end{equation}
%%%%%%
\begin{figure}[!t]% 
		\centering
 		\includegraphics[width=8.0cm,height=6.2cm,angle=0]{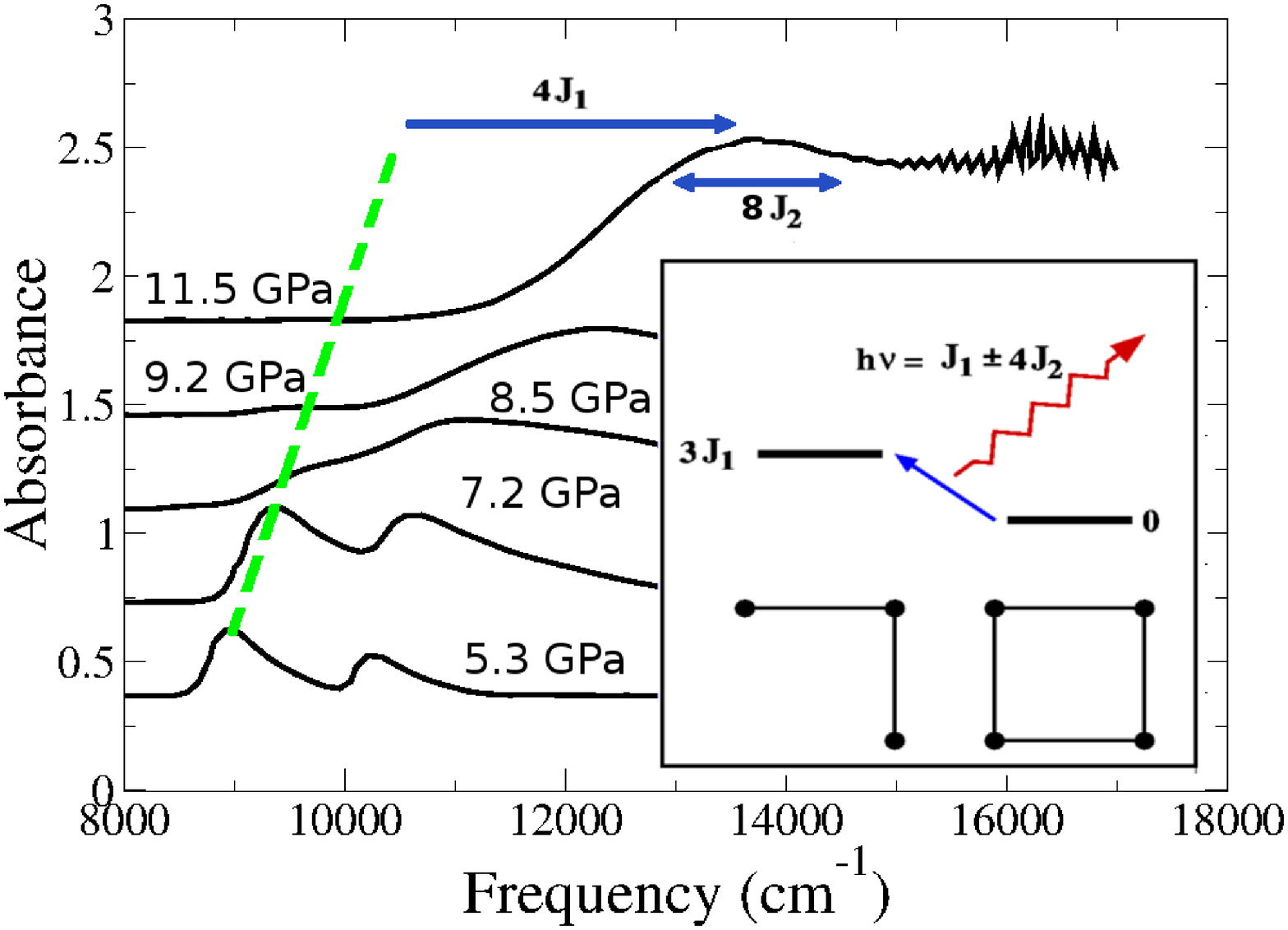}
		\caption{\label{fig:Blue.shift}{(Color online) Explanation of the blue shift at the $\delta \to \epsilon$ transition.
Solid lines are data of the IR absorption from Ref. \cite{Gorelli-PRB-2001}, dashed
green line is a linear extrapolation 
of the peak corresponding to $^3\Sigma^{-}_g\to ^1\Delta_g$ process at 11.5 GPa, the
blue arrow represents the 
blue shift of $4J_1$ with a $8J_2$ broadening represented by the double blue arrow.
Inset: Initial ``plaquette'' and final $S=1$ trimer state;  an itinerant spin-1
excitation is emitted by spin conservation (red wavy arrow) which costs an
additional energy of $h\nu \simeq 4\big(J_1\pm  J_2\big)$.}}
 \end{figure} 
 
They are not independent, since they gain an energy $-J_2/4$ when they are nearest neighbour. This  
attraction however cannot overcome the hard-core constraint, since the two triplets cannot 
reside on the same plaquette. 
The problem is thus equivalent to 
two hard-core bosons, each costing an energy $J_1$ and able to hop between nearest neighbour plaquettes with an amplitude $-J_2$. 
In spite of the weak nearest neighbour attraction $-J_2/4$, their  lowest energy state is unbound and has 
energy $2J_1-8\,J_2$. 
%%%%%%%
The overall singlet, spin liquid state will be stable so long as this excitation gap is positive, whereas 
antiferromagnetism will prevail if it is zero or negative. Based on this result, a better estimate for the stability of 
the overall singlet is the condition $2J_1 \gtrsim 8J_2$, which leads to $J_2<J_1/4$. 
The estimated values of the exchange couplings satisfy this inequality, confirming %$J_2<J_1/4$, which confirms
the stability of an overall singlet state, even once quantum fluctuations are included. 
The spin liquid singlet state, a lattice of plaquettes each made of four antiferromagnetically correlated S=1 sites, 
with weaker but nonzero inter-plaquette correlations, represents in conclusion our best model for $\epsilon_1$-O$_2$ below 20 GPa.

\subsection{Infrared spectrum in the lattice singlet model}

One burning question is at this point what evidence can one identify proving the existence of a nonzero
molecular spin in low pressure  $\epsilon$-O$_2$, despite its lack of magnetic long-range order.  
There is in fact at least one such evidence, 
long published that but not interpreted yet, and it is optical.
Near infrared spectroscopy across the $\delta$-$\epsilon$ transition shows the excitation of a single O$_2$ 
molecule from its lowest energy $^3\Sigma^{-}_g$ S=1 configuration to the lowest S=0 $^1\Delta_g$  state. 
This process, forbidden by spin and parity in isolated O$_2$, is allowed in a lattice of molecules
and shows up  in high pressure optical absorption~\cite{Gorelli-PRB-2001}. The peak corresponding to $^3\Sigma^{-}_g\to^1\Delta_g$
(and its first vibrational satellite, triggered by molecular elongation in the $S$=0 state ) is clearly visible in the $\delta$-phase, at an energy 
$\sim 8-9.000$ cm$^{-1}$, a frequency comparable to that of the isolated molecule. When the $\epsilon$-phase sets in above 8GPa the 
$^3\Sigma^{-}_g\to ^1\Delta_g$ is abruptly blue shifted  to $\sim 12.400$ cm$^{-1}$ and very considerably broadened, 
to the extent that its vibrational satellites are not anymore distinguishable. 
\cite{Gorelli-PRB-2001}  That optical observation can now be explained.

In the $S=1$ Heisenberg model representation,  the$^3\Sigma^{-}_g\to^1\Delta_g$ excitation of a single
molecule amounts to annihilating a spin 1 site in a plaquette leaving a vacancy in its place. We should expect optical absorption at a frequency equal to the 
$E_{\;^3\Sigma^{-}_g\to ^1\Delta_g}$ molecular excitation energy plus the energy cost of the molecular vacancy. 
An experimental estimate of this cost was given as a mere 250 cm$^{-1}$ in the $\delta$-phase at much lower pressure~\cite{Gorelli-PRB-2001}, 
where the absorption peak is close to the molecular excitation energy and its broadening, attributable to
shake up of spin waves, is small.  As pressure rises the exchange coupling $J_1$ increases rapidly,
as signaled by the blue shift and broadening of this transition visible within the $\delta$-phase.\cite{Gorelli-PRB-2001}  
In the  O$_2$ quartet singlet state, a vacancy costs roughly the energy difference between the ground state 
of the three surviving O$_2$ molecules and the initial four-molecule state.
This difference is readily found to be $3 J_1$.  Moreover, since the three molecule state has $S=1$, while the initial four molecule
state was a singlet, an itinerant spin-1 excitation must be created in addition to the vacancy, costing an additional energy 
$\omega\in [J_1-4J_2,J_1+4J_2]$ (see Inset in Fig.\ref{fig:Blue.shift}).  We therefore predict the absorption line to undergo at 
the $\delta \to \epsilon$ transition a sudden blue shift of $\lesssim 4\big(J_1-J_2\big)$ with a large broadening 
$\sim 8 J_2$ due to inter-plaquette exchange.  
With the calculated exchange values at 11.4 GPa that means $\simeq 0.54~\text{eV} = 4355~\text{cm}^{-1}$ 
and $\simeq 0.28~\text{eV}=2258~\text{cm}^{-1}$ respectively, values that are reasonably close to 
the experimental ones.\cite{Gorelli-PRB-2001} (see Fig.\ref{fig:Blue.shift}).  In conclusion, the abrupt
change of optical absorption is explained by the onset of a correlated
singlet state at the $\delta \to \epsilon$ transition. 
 
 \section{New Phase Diagram}

  \begin{figure}[!t]% 
		\centering
 		\includegraphics[width=8.cm,angle=0]{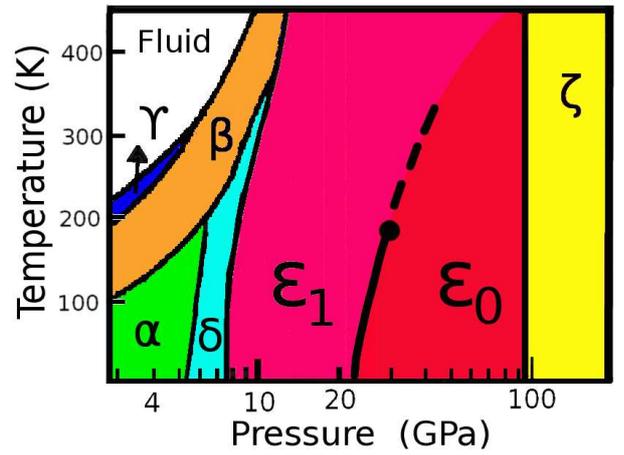}
		\caption{\label{fig:New_PDiag}{(Color online) Proposed new phase diagram of oxygen.}}
\end{figure}
%%%%%% END REVISED

The ground state of  high pressure $\epsilon$-O$_2$,  correctly described as a Peierls-distorted  nonmagnetic band insulator and an overall spin 
singlet above 20 GPa, turns below 20 GPa into a correlated insulator, where molecules recover their S=1 spin, 
but where exchange couplings and quantum fluctuations between spins conspire to yield another, 
different overall spin singlet ground state. Sharing very similar quartet lattice structures, the two states appear to have 
similar symmetry, and the question is whether the phase diagram should show a phase transition between the two 
or not. Assuming same symmetry, there could only be a first order transition or a smooth crossover. On account 
of the mechanical coupling which the onset of molecular spin must exert on the overall lattice structure, the likeliest 
candidate is a first order transition.

As it turns out, there is in literature, ignored so far by most, a rather clear evidence of a low temperature, first order
phase transition \cite{Carter.1991}, signaled around 25 GPa at 20 K by a small but sharp 
and sudden
10 cm$^{-1}$ splitting 
of a low frequency vibration. We propose that this could signal precisely the first order line separating the low pressure 
phase, say $\epsilon_1$-O$_2$, and the high pressure one, say $\epsilon_0$-O$_2$. This phase line 
will
extend at finite temperature but, on account of same symmetry, should terminate with a critical point.
Because there are underlying spins in $\epsilon_1$-O$_2$ but not in $\epsilon_0$-O$_2$, the phase line 
must 
turn towards higher pressures as temperature grows, because $\epsilon_1$-O$_2$ 
possesses a high temperature spin entropy $S \sim ln{3}$ per molecule whereas $\epsilon_0$-O$_2$
does not. Accurate room temperature vibrational data \cite{Akahama.1996}   
actually
indicates a {\it smooth} crossover between unsplit and split modes above 30 GPa,
a pressure definitely higher than the 20 K sharp transition near 25 GPa. 
That confirms our suggestion, and 
supports the prediction of a critical point below 300 K and near 30 GPa.  
Our new proposed phase diagram is therefore summarized in Fig.\ref{fig:New_PDiag}.

\section {Conclusions}

A fresh  ab initio study connects with existing vibrational evidence to indicate that molecular spin plays an important role 
in the lower pressure part of the $\epsilon$-O$_2$ oxygen phase diagram. Specifically, we propose replacing the single 
broad $\epsilon$-O$_2$ phase from 8 to 96 GPa with two phases $\epsilon_1$-O$_2$ and $\epsilon_0$-O$_2$ -- the first a local singlet spin 1
liquid, the second a regular, Peierls band insulator -- separated by a first order phase transition near 20 GPa.  The predicted 
phase line should evolve with temperature, terminating with a novel critical point, probably near 30GPa and 200 K.  The high 
temperature region below 30 GPa must be characterized by thermally fluctuating S=1 spins, whose presence should 
be directly detectable by magnetic susceptibility measurements in the 8-30 GPa pressure range. At low temperatures, the wealth of
low energy spin excitations present in $\epsilon_1- O_2$ but absent in $\epsilon_0- O_2$ should in addition give rise to very new
energy dissipation channels and processes in the former phase.

\begin{acknowledgments}
The authors would like to thank G. Baskaran, Otto Gonzales, Sadhana Chalise, Carlos Pinilla and Nicola Seriani for fruitful discussions. 
This work and in particular YC's position was partly sponsored by ERC  Advanced Grant 320796 -- MODPHYSFRICT. Contracts PRIN/
COFIN 2010LLKJBX 004 and 2010LLKJBX 007, EU-Japan Project LEMSUPER, and Sinergia CRSII2136287/1 are also acknowledged.
\end{acknowledgments}

\end{document}